# Terahertz Strong-Field Physics in Light-Emitting Diodes for Terahertz Detection and Imaging

Chen Ouyang[1,2†], Shangqing Li[1,2†], Jinglong Ma[1], Baolong Zhang[1,2], Xiaojun Wu[3*], Wenning Ren[1,2], Xuan Wang[1], Dan Wang[1,2], Zhenzhe Ma[3], Tianze Wang[1,2], Tianshu Hong[3], Peidi Yang[3], Zhe Cheng[5], Yun Zhang[5], Kuijuan Jin[1], and Yutong Li[1,2,4*]

[1]*Beijing National Laboratory for Condensed Matter Physics, Institute of Physics, Chinese Academy of Sciences, 100190 Beijing, China.*

[2]*School of Physical Sciences, University of Chinese Academy of Sciences, 100049 Beijing, China.*

[3]*School of Electronic and Information Engineering, Beihang University, 100191 Beijing, China.*

[4]*Songshan Lake Materials Laboratory, 523808 Dongguan, Guangdong, China.*

[5]*Laboratory of Solid State Optoelectronics Information Technology, Institute of Semiconductors, Chinese Academy of Sciences, 100083 Beijing, China.*

Abstract

Intense terahertz (THz) electromagnetic fields have been utilized to reveal a variety of extremely nonlinear optical effects in many materials through nonperturbative driving of elementary and collective excitations. However, such nonlinear photoresponses have not yet been discovered in light-emitting diodes (LEDs), letting alone employing them as fast, cost-effective, compact, and room-temperature-operating THz detectors and cameras. Here we report ubiquitously available LEDs exhibited gigantic and fast photovoltaic signals with excellent signal-to-noise ratios when being illuminated by THz field strengths >50 kV/cm. We also successfully demonstrated THz-LED detectors and camera prototypes. These unorthodox THz detectors exhibited high responsivities (>1 kV/W) with response time shorter than those of pyroelectric detectors by four orders of magnitude. The detection mechanism was attributed to THz-field-induced nonlinear impact ionization and Schottky contact. These findings not only help deepen our understanding of strong THz field-matter interactions but also greatly contribute to the applications of strong-field THz diagnosis.

†These authors contributed equally to this work.

*Electronic addresses: xiaojunwu@buaa.edu.cn, ytli@iphy.ac.cn

## I. INTRODUCTION

Currently, there is much interest in exploring nonperturbative nonlinear optical phenomena in various types of matter driven by intense terahertz (THz) electric and magnetic fields [1-10]. This is fueled by the impressive advancements made during the last two decades in the development of femtosecond laser-based table-top sources for strong THz radiation [11-13]. These intense THz electromagnetic fields have been utilized [14-18] to efficiently accelerate and manipulate electrons to extreme degrees, advancing the frontiers of both the technology and science of nonlinear THz-matter interactions. For example, acceleration of relativistic electron bunches in free space is promising for the realization of next-generation table-top particle accelerators and compact X-ray sources. In solids, electrons have been excited and accelerated by strong THz fields, leading to an ultrafast phase transition [19], interband tunneling and impact ionization [20], exciton generation and fluorescence [21].

These groundbreaking studies have stimulated much excitement in diverse fields of photonics, materials science, and condensed matter physics, requiring further sophistication and advancement of intense-field THz technologies. However, what is strikingly absent in this context is a convenient detector for intense THz radiation. Currently available detectors are bulky and/or expensive, often requiring liquid nitrogen or liquid helium, and are usually very slow in response. Fast, sensitive, small, inexpensive, and room-temperature-operating detectors and cameras for intense THz radiation are being sought.

Here, we report on our discovery that one of the most commonly available semiconductor devices around us, light-emitting diodes (LEDs), serve this purpose. That is, those inexpensive, ordinary devices that can even be found in children's light-up toys and warning circuits work very well for detecting and imaging THz radiation. We observed a gigantic photovoltaic signal when such LED devices were illuminated by strong-field THz pulses. This observation presents us a rare opportunity to explore the fundamental science of the ultrafast response of electron-hole pairs in a semiconductor device induced by strong-field THz pulses. More importantly, this work demonstrates that these room-temperature-operating, inexpensive LEDs are promising for developing next-generation THz detectors, cameras, and functional devices.

## II. RESULTS

**THz-field-induced photovoltaic signal in LEDs.** Intense THz pulses were generated from lithium niobate (LN) crystals via optical rectification [13,22], and a schematic diagram of the experimental setup is shown in Fig. 1a. The generated THz radiation was vertically polarized, and its strength was adjusted by a pair of THz polarizers. We focused the THz beam to achieve a fluence of, at best, ~4.1 μJ/mm$^2$; the LED detection size was only ~200×200 μm$^2$, and the focused THz beam diameter was ~3.3 mm in full width half maximum (FWHM). See more details in Supplementary Figure 1f. Under these excitation conditions, we were readily able to obtain the polarization-dependence responses via rotating the azimuthal angle of the printed circuit board (PCB) (see Fig. 1b) and monitor photosignals in an oscilloscope (Fig. 1c) in all LEDs at the same time. Typical THz temporal waveforms (see, e.g., Fig. 1d) extracted by a single-shot spectrum-encoding method [23] indicated that the generated THz pulses were near single-cycle with a frequency bandwidth of ~0.8 THz (see Supplementary Figure 1c). The measured LEDs in our experiment with different colors (red, yellow, green, blue, and white in Fig. 1e) which correspond to the bandgaps (central emission wavelength) of 1.9, 2.1, 2.3, 2.8 eV (except for the white LEDs which includes three types of LEDs of red, yellow and blue). More detailed information of the LEDs can be found in Methods.

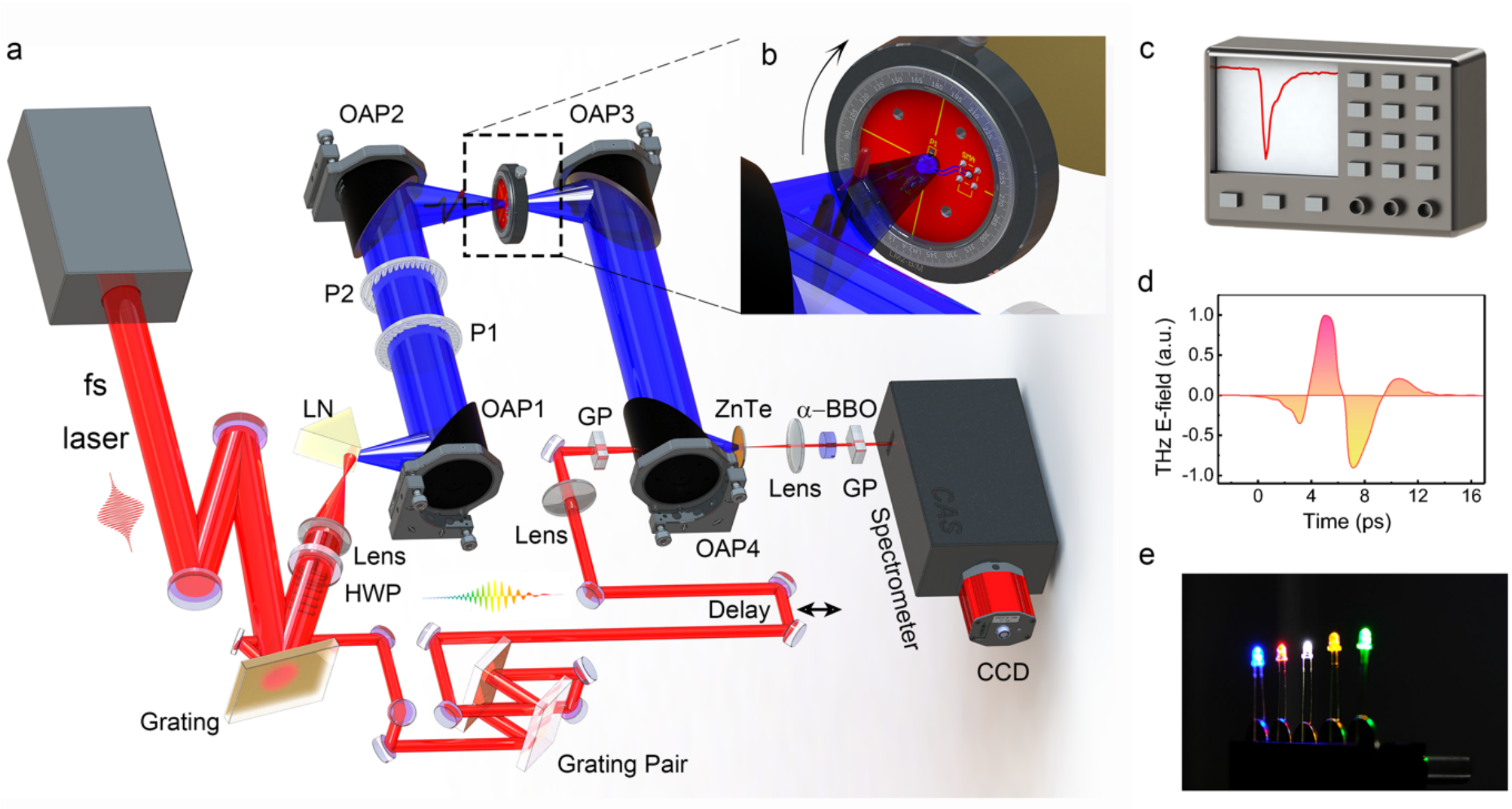

**Fig. 1. Schematic diagram of the experimental layout. a,** THz pulses for the LED detection are generated in a LN crystal via tilted pulse front technique. Two THz polarizers are used for tuning THz energies illuminated onto LEDs. The LEDs examined in our experiments include various colors of blue, white, green, yellow and red. A single-shot spectrum encoding method is employed for THz temporal wave and spectrum characterization. LEDs are positioned at the focal plane and photovoltaic signals are recorded by an oscilloscope. **b,** Zoom-in picture of LEDs which are mounted on a rotational printed circuit board for further investigate the THz polarization-dependent photoresponses. **c,** THz field induced photovoltaic signals in LEDs via impact ionization are monitored by an oscilloscope. The anode of the LED is connected to the positive electrode of the oscilloscope and vice versa. **d,** Typical THz temporal waveform recorded by the single-shot measurement. **e,** Visible light LED lamps used for THz detection. HWP: half-wave plate; LN: lithium niobate; OAP: 90° off-axis parabolic mirror; P1 and P2: THz polarizer; PCB: printed circuit board; QWP: quarter wave plate; GP: Glen-prism.

**THz field strength and polarization dependence.** To understand the origin of the photovoltaic effect in the LEDs, we examined the THz pump fluence and polarization dependence of the response of the blue LED. In the pump-fluence-dependent experiments, we used six different peak fields: 81.4, 119, 155, 190, 223, and 241 kV/cm (see Fig. 2a, the mentioned THz field strength value was the maximum peak electric field over the whole temporal duration. The details of the field strength calculation can be found in Supplementary Note 4). Observed time-dependent photoresponse curves are summarized in Fig. 2b. From this figure, we can see that all these signals are negative, i.e., the THz-pulse-induced current flew in the reverse direction, from the *n*-side to the *p*-side of the device. When the THz electric field was 81.4 kV/cm, the peak-to-peak photoresponse signal was ~41 mV, while it increased to ~600 mV when the THz field strength was increased to 241 kV/cm.

Figure 2c plots the photovoltaic signals of the LED when it was illuminated by THz radiation with three different pulse energies: 80.0, 52.8, and 24.7 μJ. The response curves exhibit a two-fold symmetric period of 180°, which is similar to the previous work on the nonlinear photoresponse of type-II Weyl semimetals [24], manifesting an anisotropic photoresponse related to crystallographic symmetry. We interpret this polarization-dependent photoresponse as a result of different threshold energy due to the orientation-dependent electron effective mass of GaN (see Methods). Figure 2d shows two photoresponse curves as a function of THz pump fluence for specific THz polarizations (marked (1) and (2) in Fig. 2c). For pump fluences lower than 1.2 μJ/mm$^2$, the response tendency exhibited a quasi-quadratic behavior, while it increased linearly when the fluence was larger than 1.2 μJ/mm$^2$. As an example, a response signal of a blue LED is illustrated in Fig. 2e. The response time of the detected photovoltaic signal was four orders of magnitude shorter than that of a commercial pyroelectric detector (SDX-1152, Gentec, sensor diameter = 8.0 mm, see Fig. 2e). Furthermore, the responsivity of these LEDs was as high as ~10 times that of the pyroelectric detector without an amplifier, reaching 750 V/W.

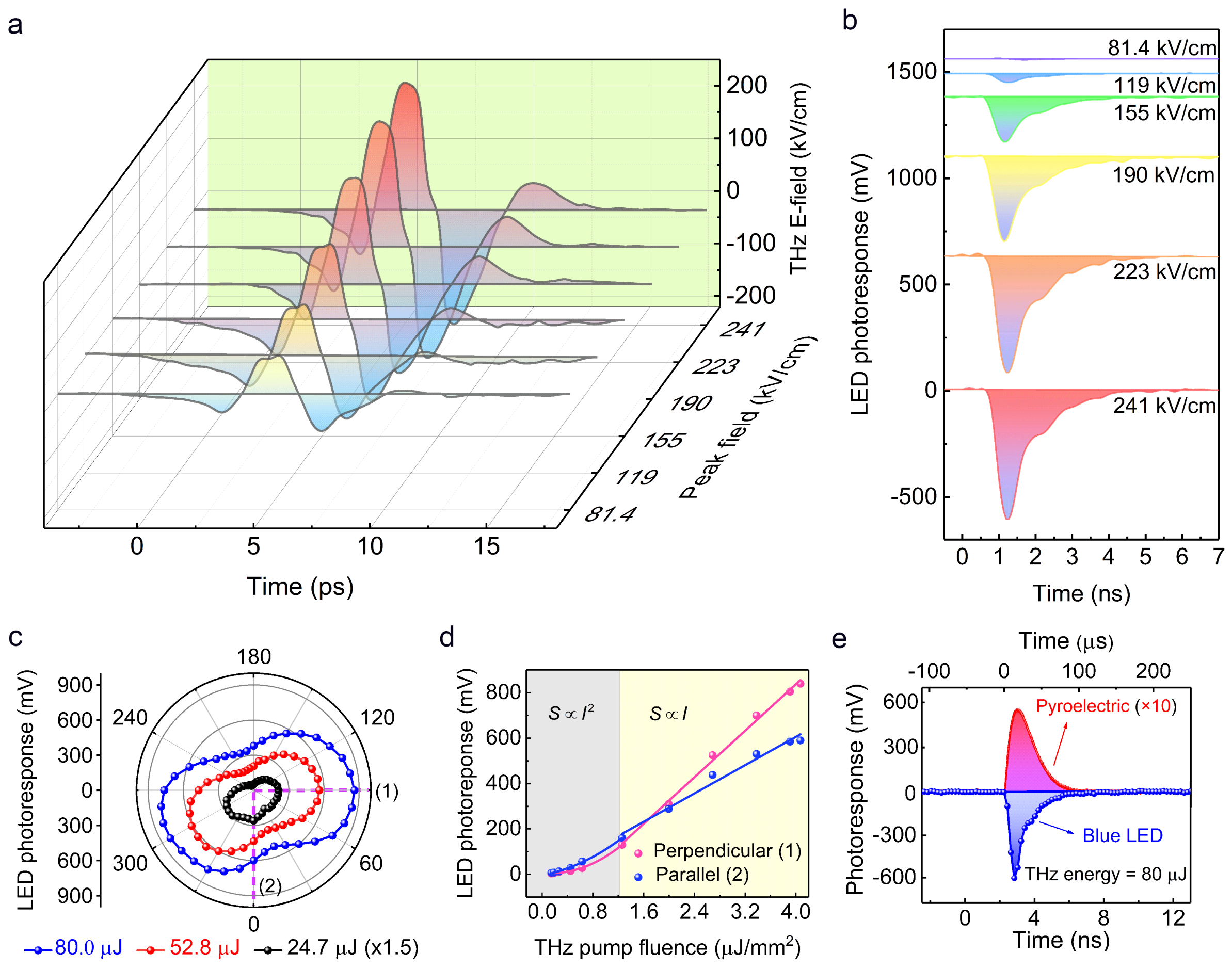

**Fig. 2 Large photovoltaic response with respect to THz field strengths. a,** Six THz time-domain waveforms obtained by tuning the THz polarizers for the blue LED detection experiments. **b.** Corresponding ultrafast photovoltaic signals monitored by the oscilloscope (load=50 Ω). The maximum photovoltaic signal is ~600 mV. **c,** Anisotropic photovoltaic signals obtained under different THz energies of 80.0, 52.8, 24.7 μJ, respectively. The photovoltaic response has strong correlation to the crystal symmetry. However, it has no strong correlation to the THz energy. Two specific THz polarization dependent responses labeled as Perpendicular (1) and Parallel (2) in **c** are systematically measured by illuminating various THz pump fluences, as exhibited in **d**. When the THz pump fluence is <1.2 μJ/mm$^2$, both polarization dependent photovoltaic signals obey quasi-quadratic relationship with respect to the THz pump fluence, while a linear behavior is predominant for the higher fluence region. **e,** Photoresponse comparison between the blue LED and the commercial pyroelectric detector

(SDX-1152, Gentec). The photoresponse signal and response time from the LED is 11 times larger and 4 orders of magnitudes faster than those obtained in the pyroelectric detector, respectively.

**THz field induced impact ionization.** When a strong THz field interacts with a semiconductor, various nonperturbative nonlinear optical phenomena can occur, including high-harmonic and sideband generation [25], the dynamic Franz-Keldysh effect [26,27], Zener tunneling [28], metallization [29], and impact ionization [6,30]. Carrier generation can result in unconventional ways through some of these processes, even though the THz photon energies used are much smaller than the bandgap. For example, Zener tunneling-induced photocarrier generation has been demonstrated, although the density achieved remained relatively low [28]; metallization can also instantaneously produce carriers but requires much higher (>100 MV/cm) field strengths. Hence, in order to explain our observed huge photovoltaic signals in the LEDs, we propose impact ionization to be the dominant mechanism. We developed a theoretical model based on this mechanism, which reproduced all salient features of our experimental results, as detailed below.

Let us first focus on the GaN-based blue LED depicted in Fig. 3a. The device structure included a sapphire substrate, a buffer layer, an *n*-type layer, activation layers, an electron-blocking layer, a *p*-type layer, and two electrodes. When the THz electric field direction is perpendicular to the stacking direction (the *z*-axis) and parallel to the *x*-*y* plane, the strength and shape of the incident THz pulse is modified through interaction with free carriers, which we take into account by scaling it down in amplitude within the multiple quantum wells (MQWs) [21]. Namely, we introduced a scaling factor of $\alpha = 1.5$ for the THz field strength. This value was obtained as the best value that fits the experimental results based on our theoretical model. It also reflected the overall screening effect of the THz temporal waveform by the induced carriers. Strictly speaking, the induced carriers should have influence on the trailing part of THz waveform, but such an averaged treatment in an overall statistical way also worked fine. From now on, the THz field strengths discussed are scaled down by this factor inside the LED.

In an impact ionization process [31], an electron in the conduction band ($e_{11}$) is subject to an external THz electric field, gains kinetic energy, and then transfers its energy to another electron in the valence band, leading to the creation of an electron-hole pair ($e_{21}$ + $h_{21}$) in addition to the initial electron which has now a reduced kinetic energy ($e_{12}$). To satisfy energy and momentum conservation, the threshold energy $E_{\mathrm{th}}$ can be written as [28]

$$E_{th} = E_g \frac{(2m_e + m_{hh})}{(m_e + m_{hh})} \tag{1}$$

where $m_{\mathrm{e}}$ (= $0.22m_0$) [32] and $m_{\mathrm{hh}}$ (= $0.85m_0$) [33] are the electron and hole effective masses, respectively, $m_0$ is the electron mass in vacuum, and $E_{\mathrm{g}}$ is the material bandgap (2.8 eV for $In_{0.12}Ga_{0.88}N$). From equation (1), we can estimate $E_{\mathrm{th}}$ to be 3.37 eV for an electron impact ionization process to occur. That is, electrons have to be accelerated to acquire 3.37 eV in order to induce impact ionization. Since a single electron-initiated impact ionization event doubles the number of electrons and creates one hole, the electron and hole densities after the system experiences impact ionization processes by $< n_I >$ times can be estimated to be $N = N_0 \times 2^{<n_I>}$ and $N_0 \times (2^{<n_I>} - 1)$, respectively, where $N_0$ is the initial electron density. Apparently, $N$ is a function of peak THz electric field $\varepsilon$, hence $N(\varepsilon)$ is implemented hereafter to represent the total carrier density at certain peak field $\varepsilon$. We neglect hole-impact processes because their large effective mass would increase the $E_{\mathrm{th}}$ to 5.02 eV, which is harder to reach.

To numerically analyze our observations, we used an equation of motion based on the Drude model:

$$\hbar \frac{dk(t)}{dt} = -q\varepsilon(t) - \hbar \frac{k(t)}{\tau} \tag{2}$$

Here, $q$ is the electron charge, $\hbar$ is the reduced Planck constant, $\varepsilon(t)$ and $k(t)$ are the THz transient electric field and the wavenumber of electron, respectively. In this equation, we added a characteristic collision time $\tau$, which prevents the accelerated electrons from transferring their energy to other electrons. The term proportional to $-k(t)/\tau$ represents the effect of phonon scattering. We set $\tau$ = 1 ps to reproduce our results. This value is larger than the reported scattering relaxation time (500 fs) [34] of the longitudinal optical (LO) phonon. Najafi *et al*. have observed that the effective time constant for a single electron-phonon coupling event can increase to a picosecond scale due to a high population density [35] or another effect like phonon bottleneck [36]. Phonon bottleneck effect can reduce the LO phonon emission and even enhance the absorption of LO phonon. The enhancement can prolong the scattering time. Therefore, in our experiment, we only need to consider the electron-LO phonon scattering behavior, and the scattering time can be prolonged up to ~ 1 ps. With $\tau$ = 1 ps and $\alpha$ = 1.5, together with a wavenumber threshold of $k_{th} = \pm 5.3 \times 10^9$ m$^{-1}$ (which was calculated based on the dispersion relation obtained with first principles calculations, see Supplementary Note 7), we found that one impact ionization event occurs (Fig. 3d) within a peak field of ~40 kV/cm (Fig. 3c) which is very close to the minimum value we can detect (~33 kV/cm inside the LED). Thus, at this noise level, the minimum field strength we can detect was ~50 kV/cm (free-space field strength). When the THz field strength is enhanced to ~160 kV/cm, the number of impact ionization events is calculated to be 9 times, under the assumption that the wavenumber is immediately reset to zero once the average wavenumber of all electrons reaches to $k_{th} = \pm 5.3 \times 10^9$ m$^{-1}$. For the given $\alpha$, $\tau$, and $k_{th}$, the obtained theoretical curve is represented as the black solid dot line in Fig. 3b.

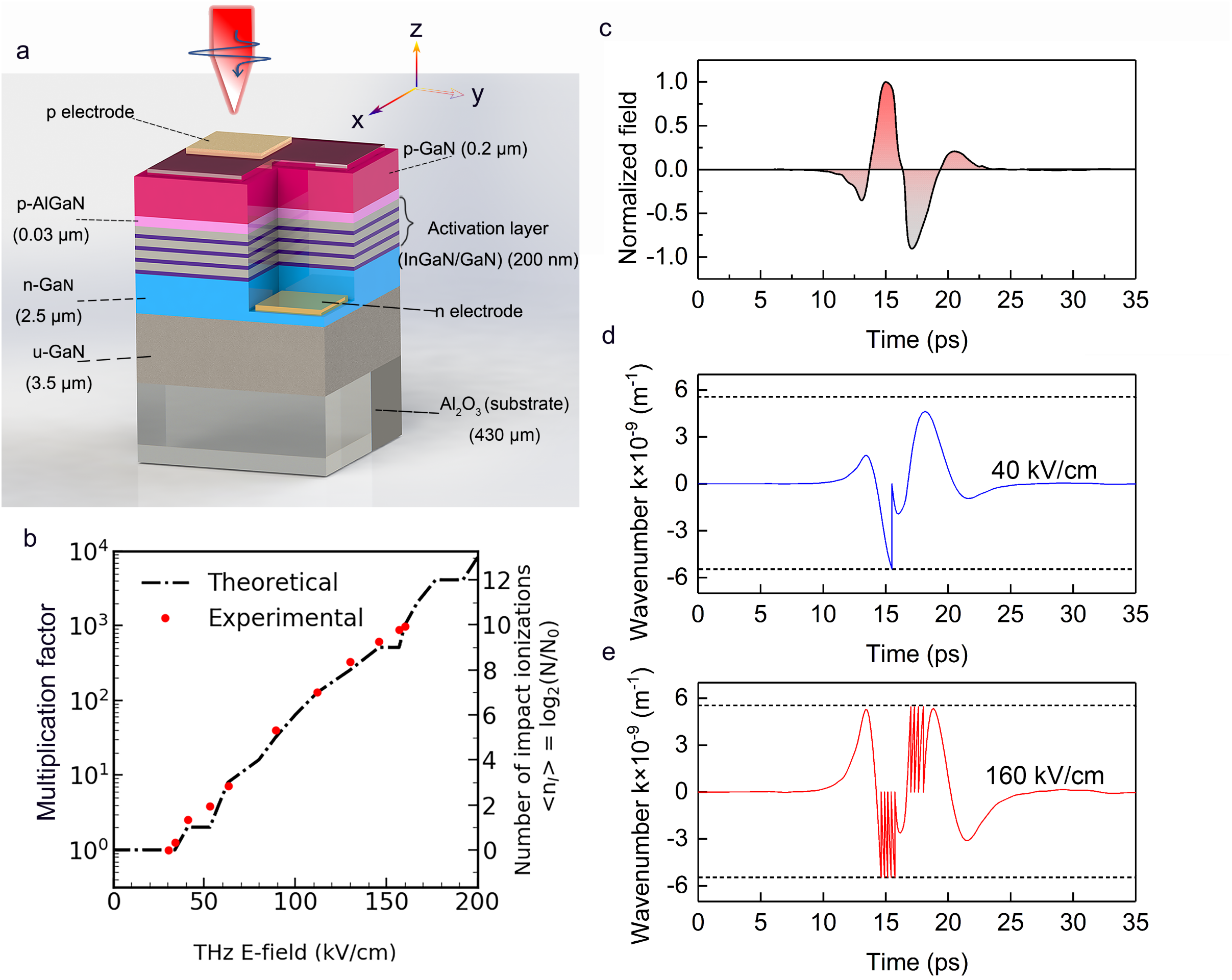

**Fig. 3 Comparison between theoretical calculation and experimental results. a,** Geometrical structure diagram of the blue LED. **b,** Photovoltaic signals determined by carrier density as well as number of impact ionizations as a function of the THz electric field strength. The theoretical calculation based on impact ionization model can well explain the experimental results. **c**, The normalized waveform of THz pulse. **d,** For the given THz electric field waveform, theoretical prediction of field strength requirement of one-time impact ionization in the blue LED is 40 kV/cm. **e,** When the THz electric field is 160 kV/cm inside the LED structure, the wavenumber of electron reaches to the wavenumber threshold (two dash lines) nine times, implying nine-time impact ionization happens. Threshold value of wavenumber $k_{th} = \pm 5.3\times 10^{9}$ m$^{-1}$ is obtained based on the dispersion relationship of $In_{0.12}Ga_{0.88}N$ which corresponds to energy threshold $E_{th} = 3.37$ eV.

In our experiments, THz photosignals were produced under open-circuit conditions (see Supplementary Figure 7). Thus, according to the Schockley-Queisser (SQ) model [37],

$$V_{OC} = \frac{n_{id} k_B T}{q} \ln(\frac{j_L}{j_0} + 1) \tag{3}$$

Where $j_L$ denotes the short-circuit photocurrent density under illumination while $j_0$ is the dark current, $q$ is the elementary charge, $k_B T$ is the thermal energy at the temperature $T$ of the LED, and $n_{id}$ is the ideality factor. If $j_L$ is proportional to THz fluence $I_0$, which is a common situation in solar cells, we should get a logarithmic relation with respect to the THz fluence. However, a linear relation severely violates the above assumption, so that in turn we can deduce an exponential relation, $j_L \propto 2^{I_0}$ (see Methods), which not only leads to a linear relation between the photovoltaic signal and THz fluence but also implies a highly nonlinear and exponentially increasing impact ionization process.

Furthermore, considering the relation of $j_L \propto N(\varepsilon)$ (see Methods), we can finally correlate the macroscopic photovoltaic signal and microscopic carrier density through the ratio

$$\frac{N(\varepsilon)}{N(\varepsilon_{\min})} \propto \frac{j_L(\varepsilon)}{j_L(\varepsilon_{\min})} \propto \frac{\exp(qV_{OC}(\varepsilon) / n_{id} k_B T - 1)}{\exp(qV_{OC}(\varepsilon_{\min}) / n_{id} k_B T - 1)} \tag{4}$$

Here, an ideality factor of 13 fits data well, implying an effective temperature of 3900 K during the whole process. Eventually, the growth of the number of impacting events $< n_I >$ due to the variation of the THz field strength from ~30 kV/cm to 160 kV/cm can be evaluated from the experimental results to be ~10 (red solid circles in Fig. 3b). Accordingly, we can also estimate the impact ionization rate to be $\gamma_I \sim 1.25 \times 10^{12}$ s$^{-1}$ within the ~8 ps THz pulse duration. We summarize all experimental data fit by the theoretical prediction in Fig. 3b, where theoretical and experimental results agree very well. This model further tells us that this linear trend will persist till the built-in field is offset by the potential difference induced by THz excited photocarriers, such that the saturation voltage will reach more than 3 V based on the built-in potential inside the blue LED. In short, we fortunately observed such gigantic photovoltaic signals in blue LEDs that stimulated much interest in investigating the impact ionization process. Although this process is common in some typical devices like photomultiplier, the large and effective photocarrier multiplication is often obscured by other effects, such as phonon absorption [38], valley scattering [39], and exciton dissociations [40]. Huge photovoltaic signals observed here show that the carrier multiplication process induced by intense THz pulses is efficient and not obscured by other effects. We believe studying this field-induced high efficiency of photocarrier multiplication is a very important driving force for future material theory physics and device optimization in the THz frequency range.

## III. THZ PHOTOVOLTAIC RESPONSE IN DIFFERENT COLOR LEDS

We next studied four other LEDs with different colors (green, white, yellow, and red, see Fig. 4a-d) and observed negative signals in all LEDs. Data were taken for $\theta = 0^\circ$ (see Supplementary Figure 3) with THz field strengths of 160, 126, and 54 kV/cm. We used Equation (3) to compare the responsivities of the four LEDs. Due to the dark current following the relation $j_0 \propto \exp(-E_g / k_B T)$, the open-circuit voltage $V_{OC}$ at a certain $j_L$ should be proportion to $E_g$. Note that the blue and green LEDs are GaN-based while the red and yellow ones are GaP-based. In fact, the threshold of the green LED is close to those of the yellow and red LEDs, so that the $j_L$ value should be close to the above situation, leading to the signal amplitude tendency $V_{OC}(green)$ > $V_{OC}(yellow)$ > $V_{OC}(red)$. Between the blue and green LEDs, $j_L$ has a greater influence on the open-circuit voltage, leading to a larger signal in the green LED. We optimized the THz photoresponse of these LEDs, and the optimal responsivity of the green LED reached 1250 V/W. Since all the materials for these LEDs are in the Ⅲ-Ⅴ group, the carrier lifetimes were roughly identical. Therefore, the decay time of all examined LEDs were in a similar time scale (~1 ns), as extracted in the experiments.

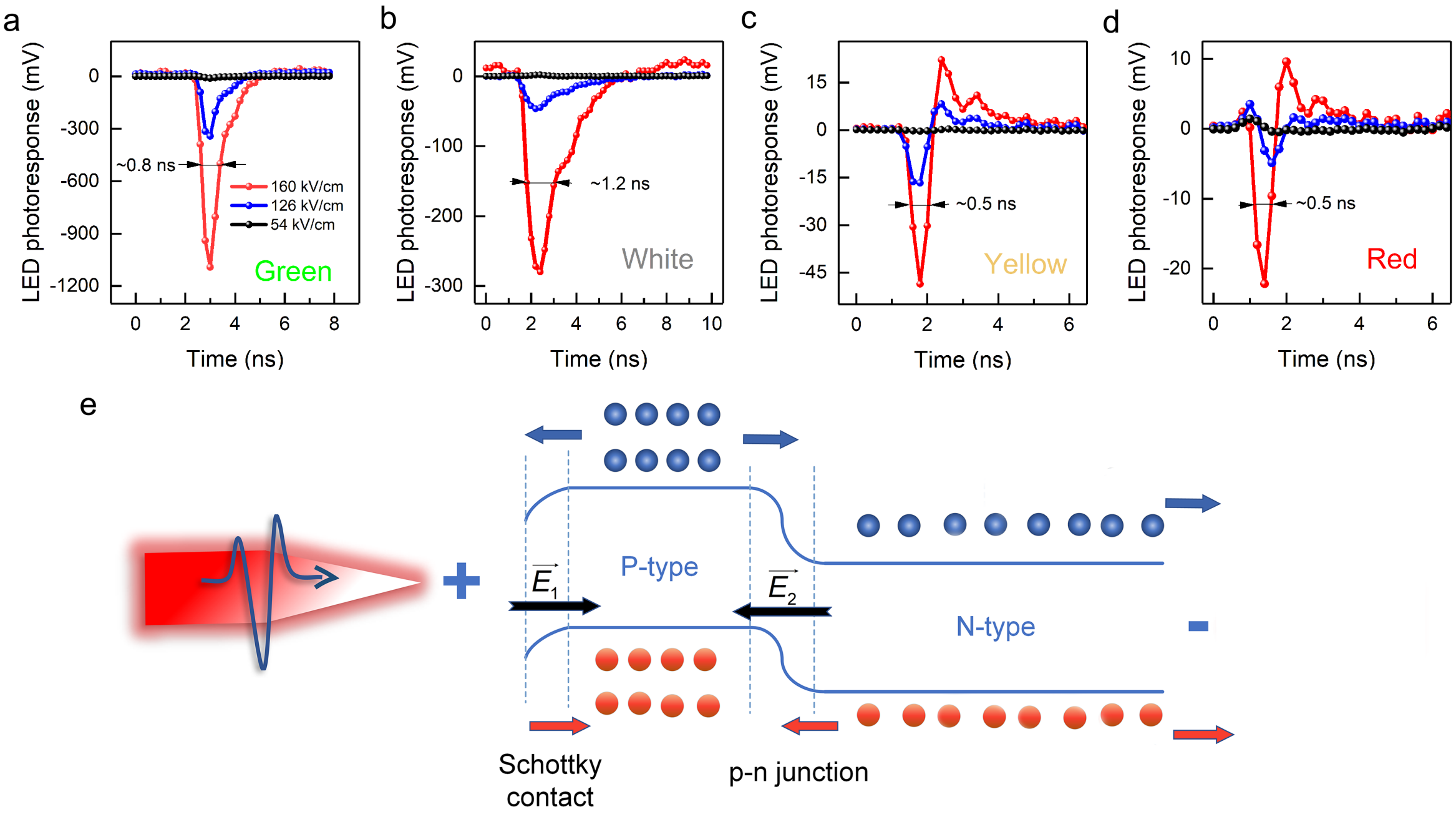

**Fig. 4 Negative photovoltaic signal analysis based on band structure. a-d**, Various photoresponse signals in LEDs (from left to right: green, white, yellow, red) under three THz electric field strength of 54, 126, 160 kV/cm, respectively. For red (**d**) and yellow (**c**) LEDs, both positive and negative photovoltaic signals are detected. **e**, Schematic diagram for explaining the simultaneous existence of positive and negative photovoltaic signals observed in yellow and red LEDs, which is attributed to the THz induced photocurrent competition between Schottky contact effect and *p-n* junction.

The most striking aspect is that the sign of the THz-photoresponse of all these LEDs is opposite to that of conventional photovoltaic signals in solar cell devices (see Fig. 2b). To explain this, we propose a Schottky contact based mechanism, as shown in Fig. 4e. Under a simple approximation, there will be a single abrupt junction between $n^+$ like metal and *p*-type semiconductor (Schottky contact), which leads to the generation of an electric field $\vec{E}_1$ with direction opposite to the electric field $\vec{E}_2$ in the *p*-*n* junction. THz induced photocarriers move towards the *n* region, and positive signals are also detectable; see Fig. 4c and d. Since the THz pulses first interact with the surface region, the induced photocarriers have a much higher density in the *p* region than in the *n* region; thus, the Schottky-based negative photovoltaic signals must be stronger than those positive signals flowing towards the *n* region. Moreover, because of the small parasitic capacitance, the negative signals are obviously faster than positive signals, further implying the existence of a Schottky contact, which fortunately improves the response time up to a few hundred picoseconds. Finally, we even observed a small positive signal in red LEDs earlier than abovementioned two signals. Among these LEDs, because the barrier height of the Schottky junction is relatively low in the red LED, hot holes within the valence band are easier to emit from the boundary of *p* side, resulting in a small positive signal. Such small positive signals among other LEDs were not observed because the Schottky barriers are higher and the hot holes cannot generate. Since this positive signal is not used in the intense THz detection, it does not affect the performance of our detectors. More experiments are needed to fully understand the intensity THz-matter interactions in such devices.

According to the proposed model in Fig. 4e, it is possible to observe a positive photovoltaic signal from LEDs if their structures did not contain Schottky junctions. We observed such a phenomenon from LEDs produced by the same company with 850 nm and 940 nm emission wavelengths. As shown in Supplementary Figure 8, appreciable positive photovoltaic signals were observed in the 850 nm LED, which can also be well understood by the proposed model. In this case, there is no Schottky junction replacing with the ohmic contact. We also observed a saturation behavior in the 940 nm LED, which has a lower saturation voltage than the 850 nm LED as expected because the saturation value is determined by the built-in potential which is proportional to the bandgap.

As a detector working in the strong-field region, LEDs have some advantages over other commercially available THz detectors. Both Golay cells and pyroelectric detectors suffer from a slow response time with ~several hundred microseconds, while bolometers usually work under low temperature conditions. Schottky barrier diodes (SBDs), which are widely used in the radio- and microwave frequency ranges, are high-speed devices (<100 ps) but require advanced material growth and device fabrication techniques. Field-effect transistors usually need a direct-current bias between the gate and source. Thus, LEDs have the advantages of high speed, broadband response, small size, low cost, no bias requirement, easy fabrication, and room-temperature operation. With regard to the Schottky contact discussed in the LEDs, it mainly affects the polarity of the photovoltaic signal and response time. Therefore, we cannot name it as a Schottky barrier diode because SBDs usually respond to a sub-cycle electric field component with a different detection mechanism. Moreover, the surface contact does not affect the THz-LED detection mechanism because we observe a positive photovoltaic signal in GaAs LEDs (see Supplementary Figure 8).

**THz-LED camera prototypes.** Since LEDs can be used for THz detection, one can straightforwardly think about fabricating a THz-LED camera. As illustrated in Fig. 5d, we demonstrated a prototype scanning THz-LED camera. We use a blue LED, and mounted it onto an automatically controlled three-dimensional translation stage with 25 μm spatial resolution. We used this prototype to scan the LED at the focal plane, obtaining the profile of a THz beam, as shown in Fig. 5a. The obtained circular beam profile has a diameter of ~2 mm (FWHM), which agrees well with that imaged by the commercial camera (Spiricon-Ⅳ, Ophire) (see Fig. 5b). LED displays and various large screen devices are widely available at low costs, indicating the feasibility of a large-area, high-resolution THz-LED real-time camera. Furthermore, we extended the THz-LED camera to a one-dimensional 1×6 array, and its prototype is given in Fig. 5e. With this LED array, we achieved a real-time THz camera, and the recorded THz beam profile is depicted in Fig. 5c. In this first-generation device, due to the large size of the market available LEDs, we could not directly image a focused THz beam. However, we could measure the THz beam profile in a non-focal plane, which in turn proves the high responsivity of the THz-LED detectors. With these results, we believe that LED-based THz detectors and cameras may have some valuable applications in strong-field THz science and technology.

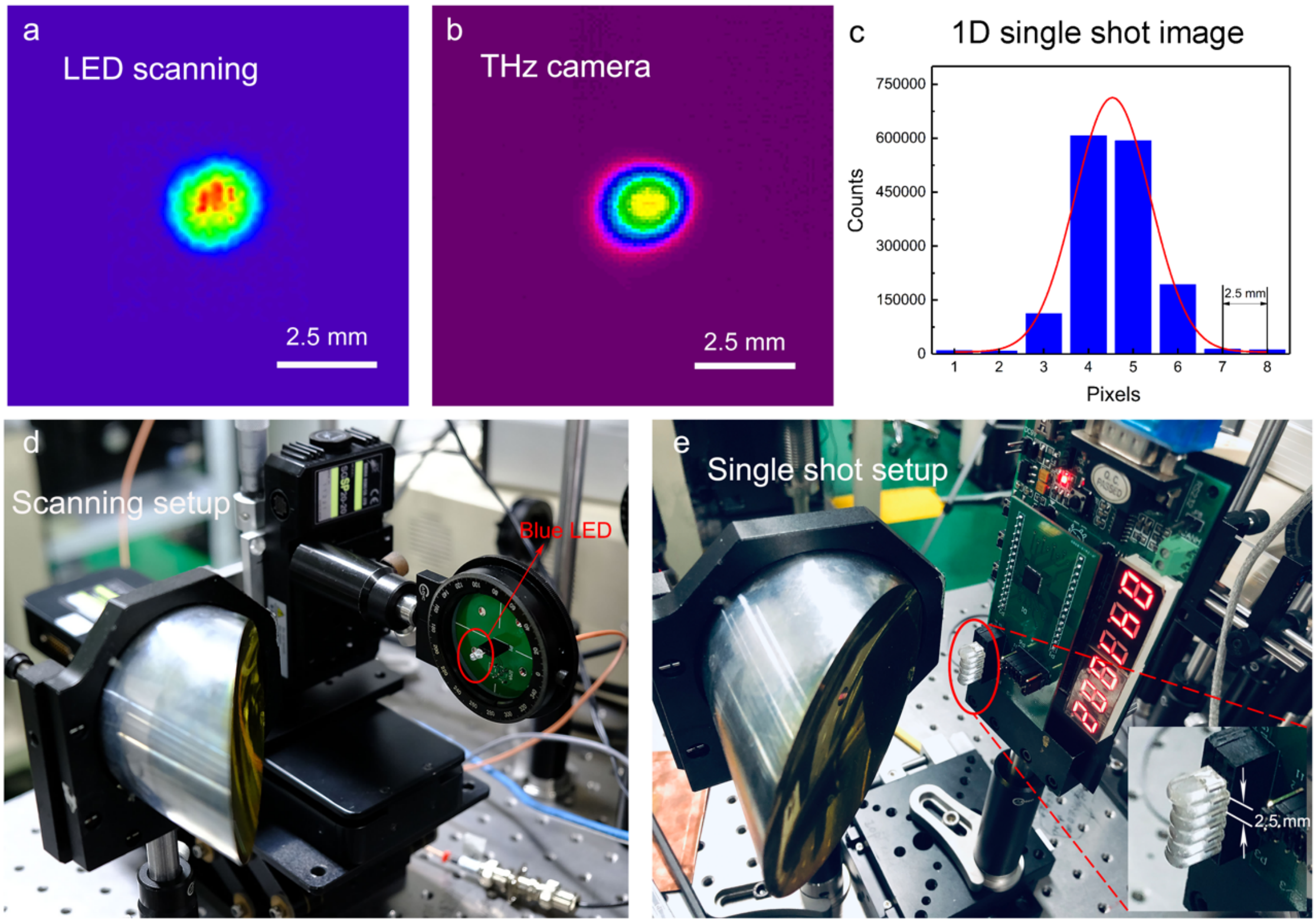

**Fig. 5 THz-LED camera prototypes and the imaged THz beam profiles. a,** The typical focused THz beam profile measured by scanning the blue LED on three-dimensional translation stages, which agrees very well with the image measured by the commercial THz camera, as shown in **b**. **c**, Measured one dimensional THz beam in the non-focal plane by the LED array, exhibiting a Gaussian-like shape (fitting result in red line). **d,** The photo for a scanning LED-THz camera prototype. **e**, The 1×6 array THz-LED camera prototype.

## IV. DISCUSSION

In summary, we showed that gigantic photovoltaic signals were produced in market available LEDs through ultrafast detection of picosecond THz transients with moderate electric field strengths. Compared with conventional photovoltaic signals, an unexpected abnormal negative signal was observed for LEDs with different colors examined in our experiments. We systematically tested several kinds of LEDs in THz polarization and pump fluence dependent studies. Experimental results can be well understood by a modified impact ionization model combined with the SQ model. Based on the high responsivity, ultrafast response time, low cost, and easy integration, we further demonstrated scanning and integrated THz-LED camera prototypes. The LED-based THz detection not only help us further deeply understand intense THz matter interaction physics, but also enables other applications such as THz real-time video imaging and coherent control of material phases.

**Acknowledgments** This work is supported by National Nature Science Foundation of China (11827807, 61905007, 11520101003, 11861121001) and the Strategic Priority Research Program of the Chinese Academy of Sciences (Grant No. XDB16010200 and XDB07030300).

**Author contributions** Y.T.L. and X.J.W. coordinate the strong-field THz radiation generation and application project. J.L.M. conceived the LED THz detection idea. C.O., S.Q.L., X.J.W., and J.L.M. carried out the experiments, collected and analyzed the data. S.Q.L. and J.L.M built the THz camera prototypes. B.L.Z., J.L.M., X.J.W. Z.Z.M. built the strong-field THz generation setup. Z.C. and Y.Z. provided some LED samples. C.O. and W.N.R. performed all simulations. B.L.Z. maintained the laser systems. X.W., D.W., T.Z.W., T.S.H., P.D.Y. and K.J.J. contributed with helpful discussions on the experimental and theoretical results. C.O., and X.J.W. wrote the manuscript with revisions by all.

**Competing financial interests** The authors declare that they have no competing financial interests.

# V. METHODS

## A. THz generation and characterization

In our experiment, a commercial Ti:sapphire laser amplifier system (Pulsar 20, Amplitude Technologies) was employed to produce strong-field THz pulses in a LN crystal based tilted pulse front technique [41]. This laser can deliver maximum single pulse energy of ~500 mJ with central wavelength of 800 nm, pulse duration of 30 fs at 10 Hz repetition rate. In order to obtain high optical-to-THz efficiency, we optimized the pulse duration by altering the group velocity dispersion of an acousto-optic programmable dispersive filter (AOPDF, Dazzler, Fastlite) [13]. In the tilted pulse front setup, a grating with density of 1480 lines/mm and of 140×140×20 $mm^3$ (length×width×height) was used for tilting the intensity of the pump pulses. A half-wave plate and a plano-convex lens with f = 85 mm focal length were inserted between the grating and the LN crystal. The LN was a z-cut congruent crystal with 6.0 mol% MgO doped to improve its damage threshold. The crystal was a triangle prism with dimensions of 68.1×68.1×64 $mm^3$ in x-y plane and the height in z-axis was 40 mm. For the STD-LED experiment, the maximum pump energy we used was ~60 mJ and the LN crystal was not cryogenically cooled, resulting in a maximum THz electric field of ~240 kV/cm with a spectrum from 0.1~0.8 THz. The radiated THz signal was first collimated by a 90° off-axis parabolic mirror (OAP1) and then went through two THz polarizers (Tydex) which were used for tuning the THz pump energy. After the second parabolic mirror (OAP2), the THz pulses were focused onto the LED which was connected to an oscilloscope (DPO 4104，Tektronix), in which a photovoltaic signal was recorded.

For THz temporal waveform and spectrum characterization, we used a single-shot spectrum coding method. As illustrated in Fig. 1a, when there was not LED in the setup, the THz waves were collimated again by OAP3 and focused together with the probing beam which came from the zero-order reflection of the grating into a 1 mm thick ZnTe detector. The probing beam was first stretched by a grating pair (Grating reticle density of 1200 lines/mm，$50\times50\times10$ mm$^3$) and then propagated through a delay line. With this method, the THz temporal electric field vector was encoded onto the chirped spectrum of the probing beam in the ZnTe crystal through electro-optic effect. The modulated probing beam spectrum was recorded by an analysis system including a lens, a BBO, a Glen-prism (GP), a spectrometer and a CCD camera, and the THz temporal waveform can be decoded and obtained. More details can be found in Supplementary Notes 1-5. The THz single pulse energy was detected by a pyroelectric detector (SDX-1152，Gentec), and its beam profile was extracted by a commercial THz camera (Pyrocam IV, Spiricon). Thus, we can estimate the error of the THz peak field strength with the information from the measured THz energy, beam size and temporal duration (see Supplementary Note 4). All these measurements were conducted at room temperature and the THz system was not purged.

### B. LED material and structure

In our experiment, the LEDs we examined were the most common LED lamps in the market for children's toys, TV sets, monitors, telephones, computers and circuit warnings. We mainly selected several kinds LEDs with different colors of blue, green, white, yellow and red for systematical investigations. The materials of blue and green LED lamps are InGaN with 3 mm round type (4204-10SUBC/C470/S400-X9-L, 04-10SUGC/S400-A5, EVERLIGHT), while the white, orange and red ones are AlGaInP (204-15/FNC2-2TVA, 204-10UYOC/S530-A3, 4204-10SURC/S530-A3, EVERLIGHT). The blue LED was exemplified, and its parameters were referred [42] in Fig. 3a. From top to bottom, the structure is composed by different layers with various thicknesses of 0.2 μm thick p-GaN, 0.03 μm thick p-AlGaN, the activation layer of 0.2 μm thick InGaN/GaN, 2.5 μm thick n-GaN, 3.5 μm thick u-GaN, and a 430 μm thick $Al_2O_3$ substrate. For this LED, it emitted blue light when a voltage of 3 V was applied. When no voltage was applied and the THz pulse illuminates onto the LED, we can probe a photovoltaic signal. We also saw similar phenomena in LEDs with other colors, but the responses were different.

### C. Sophisticated model discussion and applicability of this model

Inspection of the anisotropic curves in Fig. 2c, which is consistent with Baraff's prediction that the impact ionization process is mainly due to lucky carriers at low applied electric fields [43]. In other words, these carriers will reach threshold in some certain directions more easily than others. Thus, this phenomenon reflects that the LED certainly worked at the low-field region. Only Shockley's 'lucky carriers' [44] can be accelerated to high energy level over threshold and initiated impact ionization. Others will remain below threshold. Nonetheless, how many 'lucky carriers' are lucky enough to reach this goal is unknown. It is difficult to be solved and beyond the model in this work. Thus, in this model, we cannot evaluate the initial carrier concentration which 'triggers' the subsequent impact process. We cannot estimate the final carrier concentration generated during the THz field induced impact ionization. Only 'Multiplication factor' can be calculated out shown in Fig. 3b.

In terms of other factors which may affect the applicability of our model, we need to conduct more discussions. The assumptions of this model base on a uniform system which does not include information about defects and other channels, such as an impact ionization event occurring between a free carrier and a confined carrier within an impurity state, an impurity band or confined quantum-well state. For defects, we actually measured three regular and symmetric anisotropic two-fold photoresponse curves and plotted them in Fig. 2c. However, the defects can prevent carriers from attaining high multiplication [45] and render the photoresponse irregular while the anisotropic photoresponse of blue LEDs is relatively regular.

Therefore, we can conclude that the defects affect little and attest to the applicability of the model in this sense.

With regard to the second factor, we pick up the confined quantum-well state as an example, the impact ionization between the free carriers and confined carriers will result in the space-charge neutrality no longer being maintained [46]. This will alter the electric field profile within the device and subsequently the device performance. However, the measured Volt-ampere characteristics before and after the illumination of THz pulses exhibit no hysteresis phenomenon (see Supplementary Figure 4), implying the confined quantum-well state affect little on this model as well. This is a reasonable result because the activation layer (MQWs part) is much thinner than other layers. The confined states are very few such that the second effect plays a small role in the total carrier multiplication process.

After the carrier multiplication process, combining the previous theory with the large negative photovoltaic signal we have observed, we can infer the carrier transport process in the blue LED and green LEDs. Because the strong-field THz pulse excited carriers at 15 ps, it was quite short compared with the carrier lifetime of a few nanoseconds. The carrier multiplication can be regarded as an instantaneous process. In the blue and green LEDs, since there is no positive signal at the tail of the photovoltaic signal, it can be deduced that most of the carriers inside the device were mostly generated in the p-type region. Due to the presence of the surface electric field (Fig. 4e), the generated electrons drifted outwards. The holes diffused into the bulk. Electrons accumulated near the surface of positive electrode side, while holes accumulated in the middle of the p-type region. Subsequently, the accumulated electrons and holes formed an electric field opposite to the surface electric field direction. These fields partially canceled out the original surface electric field. In the external circuit, the electrons on the surface reduced the potential at the positive electrode, forming a large negative signal. The internal carrier recombination process gradually reduced the surface potential, and finally decayed to zero while all excess carriers recombined.

Moreover, the relaxation time $\tau$ is in fact an energy dependent parameter which decreases mostly with more energetic electrons produced when applying higher electric field strengths. Thus, this value 1 ps for $\tau$ is an average number owning statistic meaning that does not reflect the real distribution of scattering process. In the future, full quantum mechanical theoretical investigation is expected to be employed to capture more information with which advanced THz detectors will be designed and manufactured.

**D. The empirical exponential law**

In the condition of single photon absorption, namely, the common case in typical photodetectors, there is a known relation between the short circuit current $j_L$ and generated carrier density N, $j_L \propto N$. Furthermore, the generated carrier density $N$ is proportional to the pump fluence $I_0$ as well, that is $N \propto I_0$. According to the above relation, we can obtain the relation $j_L \propto I_0$ leading to the relation $V_{OC} \propto \ln I_0$, which is exactly the case in solar cell [37].

However, it should be emphasized that the experimentally obtained photovoltaic signals profoundly violated the logarithmic relation between the $V_{OC}$ and $I_0$. This behavior indicates the photoresponse in LEDs to strong-field THz pulses does not belong to the case of single photon absorption. When inspecting the equation (3), the relation $j_L \propto N$ and $N \propto I_0$, the most possible and obvious relation which can be broken down is $N \propto I_0$. The underlying impact ionization obeys exactly an exponentially increasing law. The other two relations still make sense because they are universal rules regarding the external circuits and the definition of current density rather than the mechanism in materials.

Accordingly, we have to modify the relation $N \propto I_0$ based on the results in Fig. 2d. We proposed an empirical hypothesis $N \propto 2^{I_0^\beta}$, integrated with $j_L \propto N$, resulting in $j_L \propto 2^{I_0^\beta}$ with which we can retrieved the open-circuit voltage $V_{OC} \propto \ln j_L \propto I_0^\beta \times \ln 2$. At the current stage, you can clearly and readily justify this empirical hypothesis by employing the fluence-dependent curves in Fig. 2d. In the low fluence gray region, $V_{OC} \propto I_0^2$, which gives the value of $\beta \sim 2$. In the higher fluence yellow region, $V_{OC} \propto I_0$, which gives the value $\beta \sim 1$.

It's worth noting that previous works focusing on the strong-field extreme nonlinear THz imaging also have observed exponentially increasing charges in a typical silicon made charge-coupled-device (CCD) [47], which justifies the validity of this empirical hypothesis on the other hand.

---